\documentclass[12pt,russian]{article}
 \usepackage{helvet}
 \usepackage[T1]{fontenc}
 \usepackage[cp1251]{inputenc}
 \usepackage{geometry}
 \usepackage{graphicx}
 \usepackage{setspace}
 \usepackage{babel}

 \renewcommand{\vec}[1]{\mbox{\boldmath $#1$}}

 \def\gsim{\lower.4ex\hbox{$\;\buildrel >\over{\scriptstyle\sim}\;$}}
 \def\lsim{\lower.4ex\hbox{$\;\buildrel <\over{\scriptstyle\sim}\;$}}
 
 \def\bl{\par\vskip 12pt\noindent}
 \def\bll{\par\vskip 24pt\noindent}

 \def\eT{$\eta_{_\mathrm{T}}$}

 \def\beg{\begin{eqnarray}}
 \def\ende{\end{eqnarray}}

 \def\aa{A\&A}
 \def\apj{ApJ}
 
 \def\an{Astron. Nachr.}
 
 \def\mnras{MNRAS}
 
 \def\sp{Solar Phys.}

\begin{document}

\vskip 1.0 cm

\begin{center}
{\bf Solar Dynamo Model with Nonlocal Alpha-Effect}
\end{center}

\bl

\centerline{{\bf L.~L.~Kitchatinov}$^{1,2}$\footnote{E-mail:
kit@iszf.irk.ru}, {\bf S.~V.~Olemskoy}$^1$}

\bl

\begin{center}
$^1${\it Institute for Solar–Terrestrial Physics, Russian Academy of
Sciences, Siberian Branch, P.O. Box 4026, Irkutsk,664033 Russia} \\
$^2${\it Pulkovo Astronomical Observatory, Russian Academy of
Sciences, Pulkovskoe sh. 65, St. Petersburg, 196140 Russia}
\end{center}

\bll
\hspace{0.8 truecm}
\parbox{16.4 truecm}{
{\bf Abstract}—The first results of the solar dynamo model that
allows for the diamagnetic effect of inhomogeneous turbulence and
the nonlocal alpha-effect due to the rise of magnetic loops are
discussed. The nonlocal alpha-effect is not subject to the
catastrophic quenching related to the conservation of magnetic
helicity. Given the diamagnetic pumping, the magnetic fields are
concentrated near the base of the convection zone, although the
distributed-type model covers the entire thickness of the convection
zone. The magnetic cycle period, the equatorial symmetry of the
field, its meridional drift, and the polar-to-toroidal field ratio
obtained in the model are in agreement with observations. There is
also some disagreement with observations pointing the ways of
improving the model.
\bl
{\bf DOI:} 10.1134/S1063773711040037
\bl
 }

\bl

\reversemarginpar

\setlength{\baselineskip}{0.6 truecm}

 \centerline{INTRODUCTION}
 \bl

The scale of the flows responsible for the generation of large-scale
solar magnetic fields is not small compared to the depth of the
convection zone. This primarily applies to the cyclonic flows
responsible for the so-called alpha-effect (Parker 1955; Vainshtein
et al. 1980), particularly to its emergence during the formation of
magnetic loops because of magnetic buoyancy (Caligari et al. 1995).
Active regions on the Sun are believed to emerge when deep toroidal
fields rise to the solar surface. In this case, Joy’s law holds: the
leading sunspots of active regions are closer to the equator than
the following ones (see, e.g., Obridko 1985; Howard 1996). Babcock
(1961) noted that, as a consequence of this law, the magnetic fields
of active regions must contribute to the general poloidal field of
the Sun, i.e., to the emergence of the alpha-effect. Evidence for
the presence of the alpha-effect of this type on the Sun has
recently been obtained from observations (Dasi-Espuig et al. 2010).
Since the toroidal fields are located in deep layers of the Sun and
the poloidal fields that emerge as they rise are formed near the
surface, the corresponding alpha-effect is not local (Dikpati and
Charbonneau 1999).

Figure 1 illustrates this circumstance. The nonlocal alpha-effect is
particularly interesting in connection with the problem of
catastrophic quenching of the alpha-effect (Brandenburg and
Subra-manian 2005). Basically, the problem consists in the
following. The generation of a poloidal field because of the
alpha-effect gives rise to a large-scale field with magnetic
helicity. In view of the conservation of helicity, helicity equal in
magnitude and opposite in sign must appear in a small-scale magnetic
field. In turn, the helical small-scale field contributes to the
alpha-effect that is opposite in sign to the already existing
alpha-effect. As a result, as the large-scale field grows, the
alpha-effect disappears and the field generation stops. However,
this takes place only for the local alpha-effect. If the original
toroidal field and the generated poloidal field are separated in
space, then no magnetic helicity emerges and there is no
catastrophic quenching (see, however, Brandenburg and K\"apyl\"a
2007). In Fig. 1, there is no linkage of the magnetic flux tubes of
the poloidal field and the original toroidal field that must be
present for helical magnetic structures (Moffat 1978). In this
paper, we propose a solar dynamo model with a nonlocal alpha-effect.

\begin{figure}[htb]
 \centerline{
 \includegraphics[width=12cm]{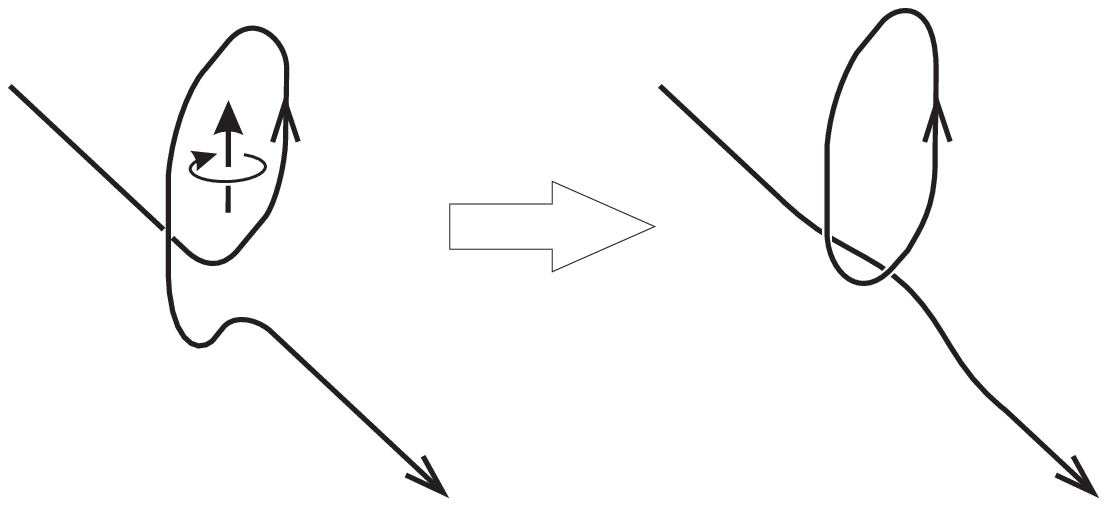}}
 \begin{description}
 \item{\small {\bf Fig. 1.} Illustration of the emergence of the alpha-effect
    due to the rise of magnetic flux tubes. Under the action of the
    Coriolis force, the rising magnetic loop acquires a poloidal
    orientation. Reconnection at the loop footpoints disconnects it from
    the original magnetic flux tube of the toroidal field. The resulting
    toroidal and poloidal loops have no linkage and possess no magnetic
    helicity.
    }
 \end{description}
\end{figure}

Another peculiarity of the proposed model is the concentration of
the toroidal magnetic field near the base of the convection zone.
This region is believed to be preferable for dynamo action (see,
e.g., Gilman 1992). The dynamo models for the near-bottom region of
the convection zone reproduce many features of the solar cycle
(Parker 1993; Mason et al. 2008; Nefedov and Sokoloff 2010).
However, the reasons why the magnetic field is concentrated in this
region have not been specified. If the distributed-type model covers
the entire thickness of the convection zone, then the magnetic field
also turns out to be distributed over the entire thickness (Dikpati
and Charbonneau 1999). In this paper, we take into account the
diamagnetic effect of inhomogeneous turbulence, which leads to the
field concentration near the bottom of the convection zone.

Turbulent conducting fluids expel large-scale magnetic fields into
regions of space with a relatively low turbulence intensity
(Zel’dovich 1956), i.e., turbulent conducting fluids possess
diamagnetic properties with respect to large-scale fields. The
presence of fluctuating magnetic fields had long been thought to
suppress the diamagnetic effect of turbulence. However, both
numerical (Dorch and Nordlund 2001; Brandenburg et al. 2010) and
laboratory (Spence et al. 2007) experiments confirm its reality. The
importance of this effect for the solar dynamo was pointed out in a
number of papers (Ivanova and Ruzmaikin 1976; Krivodubskii 1984;
R\"udiger and Brandenburg 1995; Guerrero and de Gouveia Dal Pino
2008). It is also important for the formation of the solar
tachocline (Kitchatinov and R\"udiger 2008). Allowance for the
diamagnetic pumping of the field in the dynamo model considered here
provides the concentration of magnetic fields near the bottom of the
convection zone.

The observed toroidal-to-poloidal field ratio suggests the presence
of such a concentration on the Sun. The poloidal (polar) field has a
strength of about 1–2 G. The toroidal fields are approximately a
factor of 1000 stronger if their strength is judged from the sunspot
fields. The relative latitudinal inhomogeneity of the angular
velocity is about 30\%. Such rotation inhomogeneity can produce a
toroidal field that is approximately a factor of 40 stronger than
the existing poloidal field in 11 years (strong rotation
inhomogeneity in the tachocline does not change this estimate,
because the radial field component here must be smaller than the
meridional one by the same factor by which the radial rotation
inhomogeneity is larger than its latitudinal inhomogeneity).
Therefore, the presence of poloidal fields with a strength of $\sim$
100 G is needed for the differential rotation to be able to produce
toroidal fields with a strength of several thousand gauss during the
solar cycle. The magnetic field concentration near the bottom of the
convection zone in the proposed model provides the required poloidal
field strength.

In this paper, we discuss the first results of the dynamo model that
simultaneously allows for the diamagnetic pumping of the field and
the nonlocal alpha-effect. In a number of parameters, the results
show agreement with observations.

 \bll
 \centerline{THE MODEL}
 \bl

The velocity of diamagnetic pumping of the field is related to the
inhomogeneity of turbulent magnetic diffusivity \eT~(Krause and
R\"adler 1980) by

 \begin{equation}
    {\vec U}_\mathrm{dia} =
    -\frac{1}{2}{\vec\nabla}\eta_{_\mathrm{T}}.
    \label{1}
 \end{equation}
Given this pumping, the induction equation for large-scale fields
can be written as
 \begin{equation}
    \frac{\partial{\vec B}}{\partial t} = \mathrm{curl}
    \left({\vec V}\times{\vec B} - \sqrt{\eta_{_\mathrm{T}}}
    \mathrm{curl}(\sqrt{\eta_{_\mathrm{T}}}{\vec B})
    + {\vec{\cal A}}\right) ,
    \label{2}
 \end{equation}
where $\vec{\cal A}$ allows for the contribution of the
alpha-effect. In nonlocal formulation, it can be written as
(Brandenburg et al. 2008)
 \begin{equation}
    \vec{\cal A}({\vec r}) = \int \alpha({\vec r},{\vec r}')
    {\vec B}({\vec r}')\ \mathrm{d}^3r'.
    \label{3}
 \end{equation}
We will refine the form of the function $\alpha ({\vec r},{\vec
r}')$ below. The fluid velocity $\vec V$ in (\ref{2}) corresponds to
(inhomogeneous) rotation,
 \begin{equation}
    {\vec V} = {\vec e}_\phi r\sin\theta\Omega f(r/R_\odot,\theta ) ,
    \label{4}
 \end{equation}
where the ordinary spherical coordinates ($r,\theta,\phi$) are used,
$\Omega$ is the mean angular velocity, ${\vec e}_\phi$ is a unit
vector in the azimuthal direction, $f(x,\theta )$ is the
dimensionless rotation frequency, and $x = r/R_\odot$ is the
relative radius. We will use the approximation of the
helioseismological data on the Sun’s inhomogeneous rotation proposed
by Belvedere et al. (2000)
 \begin{equation}
    f(x,\theta) = \frac{1}{461}\sum\limits_{m=0}^{2} \cos\left( 2m\left(\frac{\pi}{2}
    - \theta\right)\right) \sum\limits_{n=0}^{4} C_{nm}x^n.
    \label{5}
 \end{equation}
The values of the coefficients $C_{nm}$ are given in Table 1 from
Belvedere et al. (2000). Figure 2 shows angular velocity isolines
for the rotation specified in this way.

\begin{figure}[htb]
 \centerline{
 \includegraphics[width=9cm]{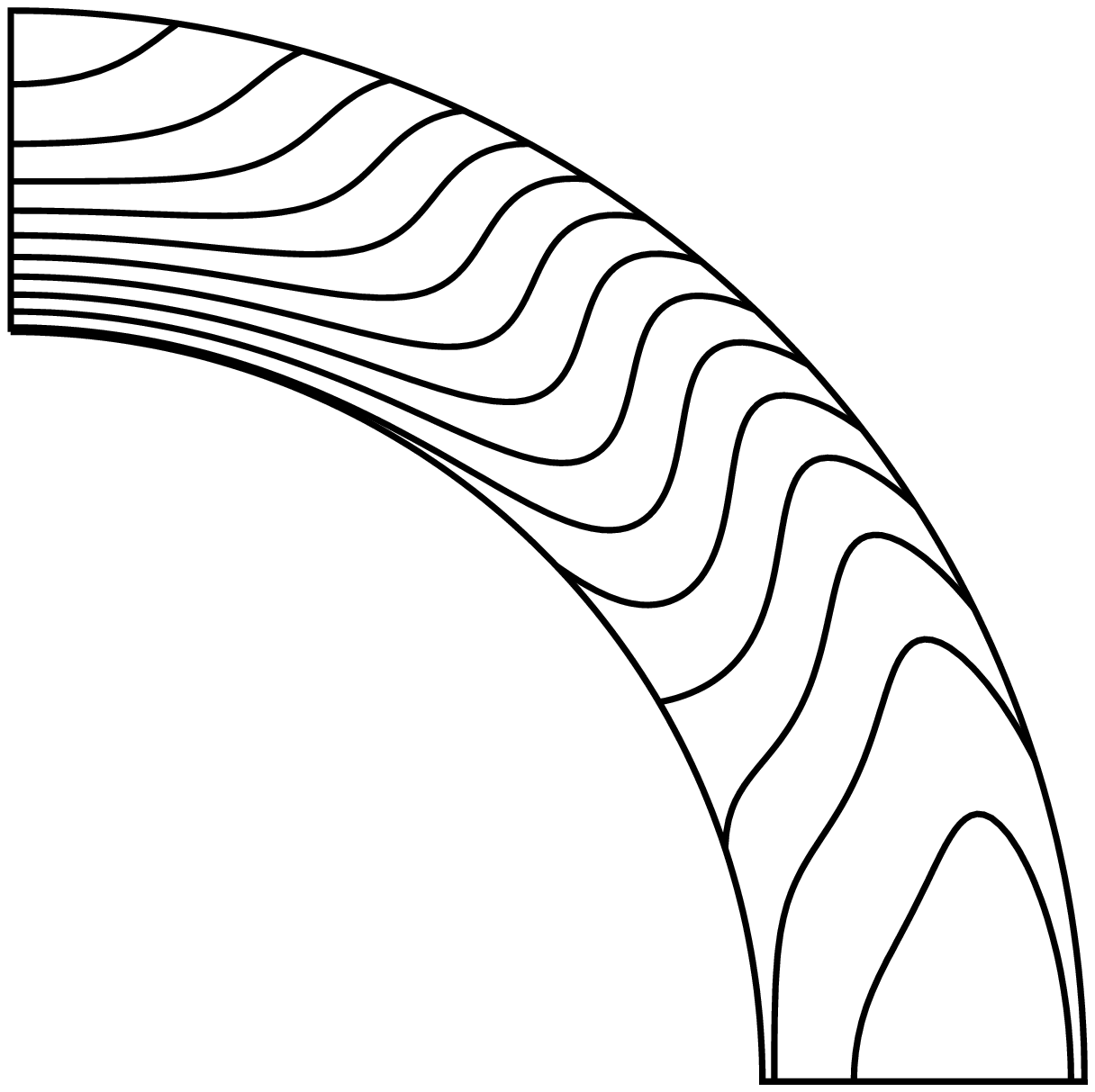}}
 \begin{description}
 \item{\small {\bf Fig.~2.} Angular velocity isolines for the differential rotation
    used in the model. The angular velocity distribution is the
    approximation of the helioseismological data on the internal solar
    rotation (Belvedere et al. 2000).
    }
 \end{description}
\end{figure}

Let us consider a two-dimensional dynamo, i.e., we will assume the
magnetic field to be cylindrically symmetric relative to the
rotation axis:
 \begin{equation}
   {\vec B} = {\vec e}_\phi B + \mathrm{curl}\left({\vec
   e}_\phi\frac{A}{r\sin\theta}\right) ,
   \label{6}
 \end{equation}
where $B$ is the toroidal magnetic field and $A$ is the potential of
the poloidal field. We will apply the $\alpha\Omega$- dynamo
approximation in which the generation of a toroidal field due to the
alpha-effect is neglected (the toroidal field is generated by the
differential rotation much more efficiently).

We use dimensionless variables: the time is measured in units of
$R^2_\odot/\eta_0$ ($\eta_0$ is the characteristic turbulent
diffusivity in the convection zone), the magnetic field is in units
of the field strength $B_0$ for which the nonlinear effects are
important, the function $\alpha$ is in units of $\alpha_0$ (its
characteristic amplitude), and the potential of the poloidal field
is in units of $\alpha_0 B_0 R_\odot^3/\eta_0$. We will retain the
previous notation for all of the normalized quantities, except for
the relative radius $x = r/R_\odot$ and diffusivity $\eta =
\eta_{_\mathrm{T}}/\eta_0$. The equation for the toroidal field in
dimensionless variables will be written as
 \begin{eqnarray}
    \frac{\partial B}{\partial t} &=& \frac{\cal D}{x}
    \left(\frac{\partial f}{\partial x}\frac{\partial
    A}{\partial\theta} - \frac{\partial f}{\partial\theta}
    \frac{\partial A}{\partial x}\right)
    \nonumber \\[0.1 truecm]
    &+& \frac{\eta}{x^2}\frac{\partial}{\partial\theta}\left(
    \frac{1}{\sin\theta}\frac{\partial(\sin\theta
    B)}{\partial\theta}\right) +\frac{1}{x}\frac{\partial}{\partial
    x}\left(\sqrt{\eta}\ \frac{\partial(\sqrt{\eta}\ xB)}
    {\partial x}\right) ,
 \label{7}
 \end{eqnarray}
where
 \begin{equation}
    {\cal D} = \frac{\alpha_0 \Omega R_\odot^3}{\eta_0^2}\
 \label{8}
 \end{equation}
is the dynamo number that is the main controlling parameter of the
model. We will write the equation for the poloidal field with the
nonlocal alpha-effect as
 \begin{eqnarray}
    \frac{\partial A}{\partial t} &=&
    \frac{\eta}{x^2}\sin\theta\frac{\partial}{\partial\theta}
    \left(\frac{1}{\sin\theta}\frac{\partial
    A}{\partial\theta}\right) + \sqrt{\eta}\frac{\partial}{\partial
    x} \left(\sqrt{\eta}\frac{\partial A}{\partial x}\right)
    \nonumber \\
    &+& x \sin\theta  \cos\theta \int\limits_{x_\mathrm{i}}^x
    \alpha (x,x') B(x',\theta)\ \mathrm{d} x' ,
 \label{9}
 \end{eqnarray}
where the integration is only over the radius and is bounded by the
upper limit $x$. This qualitatively reflects the fact that the
magnetic fields rising from deeper regions $x'< x$, contribute to
the alpha-effect at some point $x$ and the velocities of the rise
deviate only slightly from the vertical direction. The alpha-effect
due to the rise of magnetic loops was discussed in many papers, but
the available results are not sufficient for its unambiguous
quantitative allowance. It is necessary to invoke qualitative
considerations. We will write the function $\alpha (x,x')$ from
(\ref{9}) as
 \begin{eqnarray}
    \alpha (x,x') &=& \frac{\phi_\mathrm{b}(x')\phi_\alpha (x)} {1 + B^2(x',\theta)}
    ,
    \nonumber \\
    \phi_\mathrm{b}(x') &=& \frac{1}{2}\left( 1 -
    \mathrm{erf}\left( (x' - x_\mathrm{i}-2.5h_\mathrm{b})/h_\mathrm{b}\right)\right) ,
    \nonumber \\
    \phi_\alpha (x) &=& \frac{1}{2}\left( 1 +
    \mathrm{erf}\left( (x - 1 + 2.5h_\alpha)/h_\alpha\right)\right),
    \label{10}
 \end{eqnarray}
where $x_\mathrm{i}$ is the radius of the inner boundary of the
distributed-type model, $h_\mathrm{b}$ and $h_\alpha$ are the
numerical parameters of the model, and erf is the error function.
The function $\phi_\mathrm{b}(x')$ defines the (near-bottom) region
whose toroidal fields produce the alpha-effect and the function
$\phi_\alpha (x)$ describes the near-surface region in which this
effect emerges. The quantity $1 + B^2$ in the denominator of
(\ref{10}) allows for the nonlinear suppression of the alpha-effect.
Strong fields rise rapidly and the transformation of toroidal fields
into poloidal ones due to the Coriolis force becomes inefficient.
Similar views of the alpha-effect were used by Durney (1995),
Dikpati and Charbonneau (1999), and Catterjee et al. (2004). The
results discussed below were obtained for $x_\mathrm{i} = 0.7$,
$h_\mathrm{b} = 0.002$, and $h_\alpha = 0.02$.

\begin{figure}[htb]
 \centerline{
 \includegraphics[width=9cm]{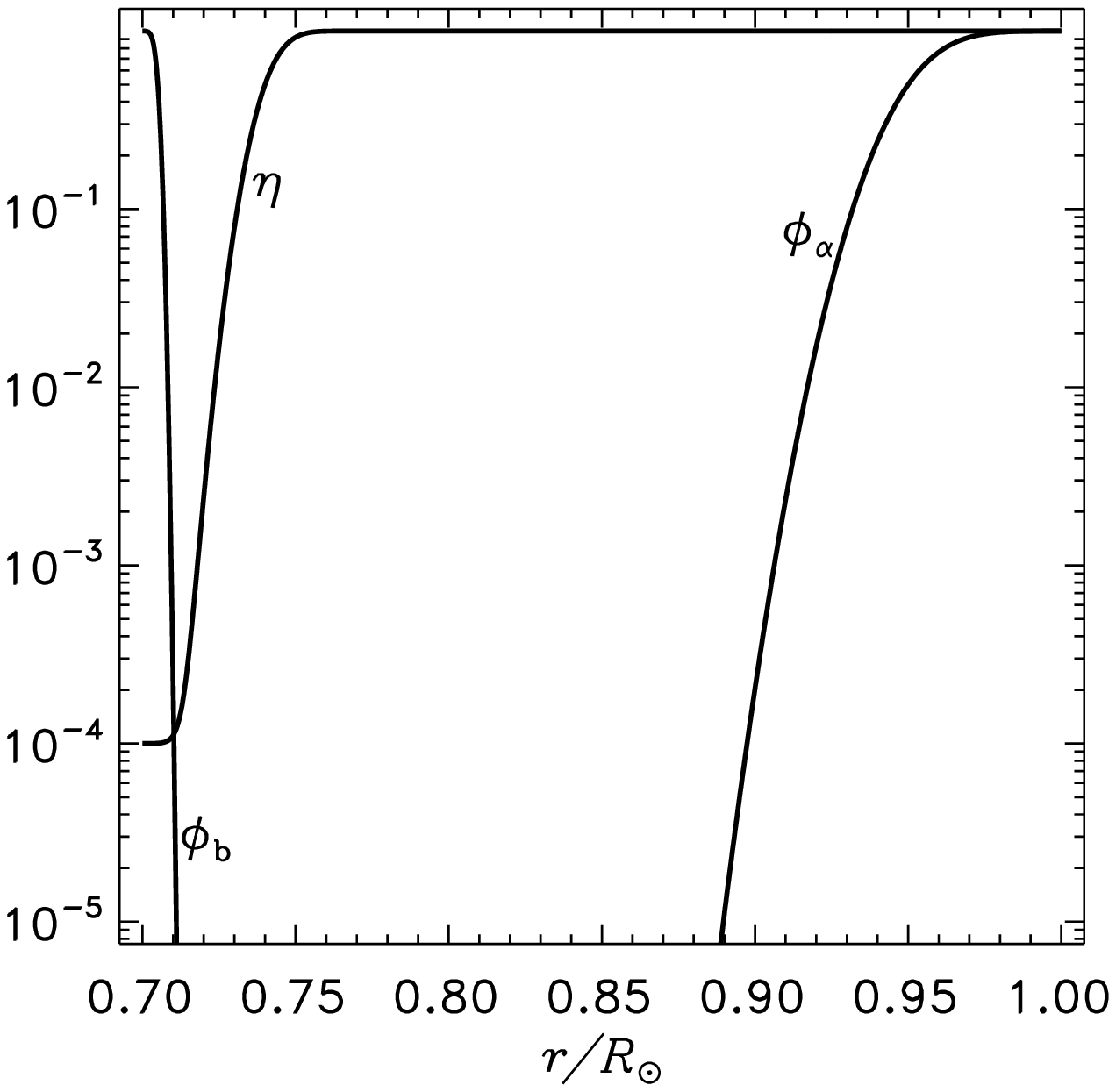}}
 \begin{description}
 \item{\small {\bf Fig.~3.} Magnetic diffusivity profile, along with
    the functions $\phi_\mathrm{b}$ and $\phi_\alpha$ defining
    the nonlocal alpha-effect.
    }
 \end{description}
\end{figure}

We use the following magnetic diffusivity distribution:
 \begin{equation}
    \eta (x) = \eta_\mathrm{in}  + \frac{1}{2}\left( 1 -
    \eta_\mathrm{in}\right) \left( 1 +
    \mathrm{erf}\left(\frac{x-x_\eta}{h_\eta}\right)\right) .
 \label{11}
 \end{equation}
Figure 3 shows the magnetic diffusivity profile for the parameters
adopted in our calculations ($\eta_\mathrm{in} = 10^{-4}$, $x_\eta =
0.74$, $h_\eta = 0.01$), along with the functions $\phi_\mathrm{b}$
and $\phi_\alpha$ from Eq. (\ref{10}) that define the nonlocal
alpha-effect.

The conditions at the lower boundary correspond to the interface
with a superconductor,
 \begin{equation}
   \frac{\partial \left( \sqrt{\eta} x B\right)}{\partial x} = 0,\ \
    A = 0,\ \ x = x_\mathrm{i} .
    \label{12}
 \end{equation}
At the upper boundary, only the radial field component is nonzero
(pseudo-vacuum boundary conditions):
 \begin{equation}
    B = 0,\ \ \frac{\partial A}{\partial x} = 0,\ \ x = 1.
    \label{13}
 \end{equation}
The initial-value problem for the system of equations (\ref{7}) and
(\ref{9}) was solved numerically. Since allowance for the
diamagnetic pumping of the field leads to strong inhomogeneity of
the solutions near the base of the convection zone, we used a
radially nonuniform (and latitudinally uniform) grid. The grid
spacing was $\Delta x \sim \sqrt{\eta}$. All of the results
discussed below do not depend on the numerical resolution (this was
checked by the repetition of calculations with doubled resolution).

Two types of symmetry of the solutions relative the equator are
possible in linear dynamo problems: antisymmetric solutions, or
dipolar modes, for which $B$ and $\partial A/\partial\theta$ are
antisymmetric relative to the equator ($B(\theta) = - B(\pi
-\theta)$), and symmetric solutions, or quadrupolar modes
(($B(\theta) = B(\pi -\theta)$). In nonlinear dynamo models (to
which our model belongs), the solutions may not have a certain
equatorial symmetry. Nevertheless, the symmetry of the solutions can
be specified by additional boundary conditions on the equator. In
most calculations, no equatorial symmetry was prescribed, but the
solutions approached a certain type of symmetry after a sufficiently
long calculation time. When it was required to determine the
conditions for the excitation of symmetric or antisymmetric magnetic
field modes, we applied the boundary conditions on the equator.

 \bll
 \centerline{RESULTS AND DISCUSSION}
 \bl

The critical value of the dynamo number (\ref{8}) for our model is
${\cal D}_\mathrm{c} = 1.9\times 10^4$. The field generation takes
place for ${\cal D} > {\cal D}_\mathrm{c}$. Fields antisymmetric
relative to the equator are generated. The global solar magnetic
field also belongs to this type of symmetry. To be more precise,
only the antisymmetric component of the fields observed on the Sun
exhibits an 11-year cyclicity (Stenflo 1988). The symmetric
component is characterized by approximately a factor of 10 lower
amplitude and by irregular variations with time. The critical dynamo
number for the generation of symmetric modes (${\cal D}_\mathrm{q} =
2.5\times 10^4$) is considerably larger than ${\cal D}_\mathrm{c}$.
The so obvious preference of dipolar modes is characteristic of
dynamo models with a relatively low magnetic diffusivity near the
base of the convection zone (Chatterjee et al. 2004).

\begin{figure}[htb]
 \centerline{
 \includegraphics[width=12cm]{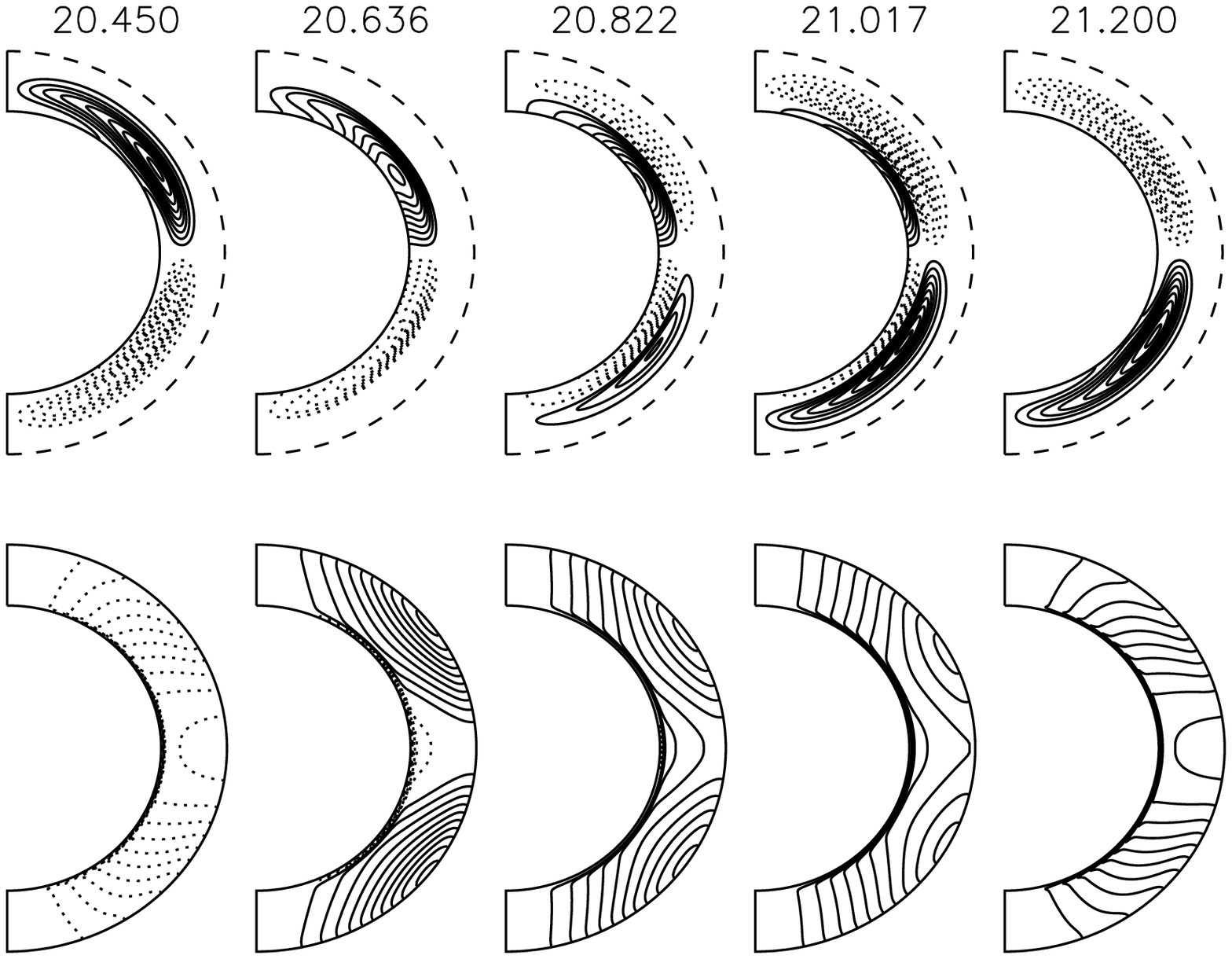}}
 \begin{description}
 \item{\small {\bf Fig.~4.} Toroidal field isolines (top) and poloidal
      field lines (bottom) for several instants of one magnetic
      cycle. The time in dimensionless units ($R_\odot^2/\eta_0$)
      is shown in the upper part of the figure. The solid lines
      indicate positive levels and field lines with clockwise
      circulation, while the dotted lines indicate negative
      levels and counterclockwise circulation. In the upper row
      of graphs, the scale along the radius was changed in such
      a way that the dashed boundary corresponds to the radius
      $r = 0.74R_\odot$ below which the toroidal field is localized.
      The results of our calculations for ${\cal D} = 2.2\times 10^4$
      are shown.
      }
 \end{description}
\end{figure}

In all likelihood, the reason is as follows. For modes with a
dipolar equatorial symmetry, the spatial scale of the poloidal field
(in latitude) is larger than that of the toroidal one. For
quadrupole modes, the toroidal field has a larger scale. The
poloidal field is distributed over the entire thickness of the
convection zone, while the toroidal field is concentrated near its
base, where the magnetic diffusion is small (Fig. 4). Therefore, the
poloidal field is subject to the destructive action of magnetic
diffusion to a greater extent and the (antisymmetric) modes whose
poloidal field has a larger scale are generated more easily.

Figure 4 shows the field distribution in the convection zone and its
variations in the magnetic cycle. The strongest fields are
concentrated near the bottom of the convection zone. In the model
under consideration, such a concentration results from the
diamagnetic pumping of the field (Zel’dovich 1956). However, the
pumping does not affect the radial field component parallel to the
pumping direction. Therefore, the poloidal field emerges on the
outer surface. Nevertheless, the poloidal field reaches its largest
strength near the lower boundary. That is why the differential
rotation can generate fairly strong toroidal fields during the
magnetic cycle. The polar field on the surface is approximately a
factor of 1000 weaker than the characteristic toroidal field in our
model. The period of the activity cycle (half the period of the full
magnetic cycle) for Fig. 4 is about $0.75R_\odot^2/\eta_0$; for
$\eta_0\simeq 10^{13}$~cm$^2 $~s.$^{-1}$, this gives a value close
to the 11-year period of the solar cycle.

\begin{figure}[htb]
 \centerline{
 \includegraphics[width=12cm]{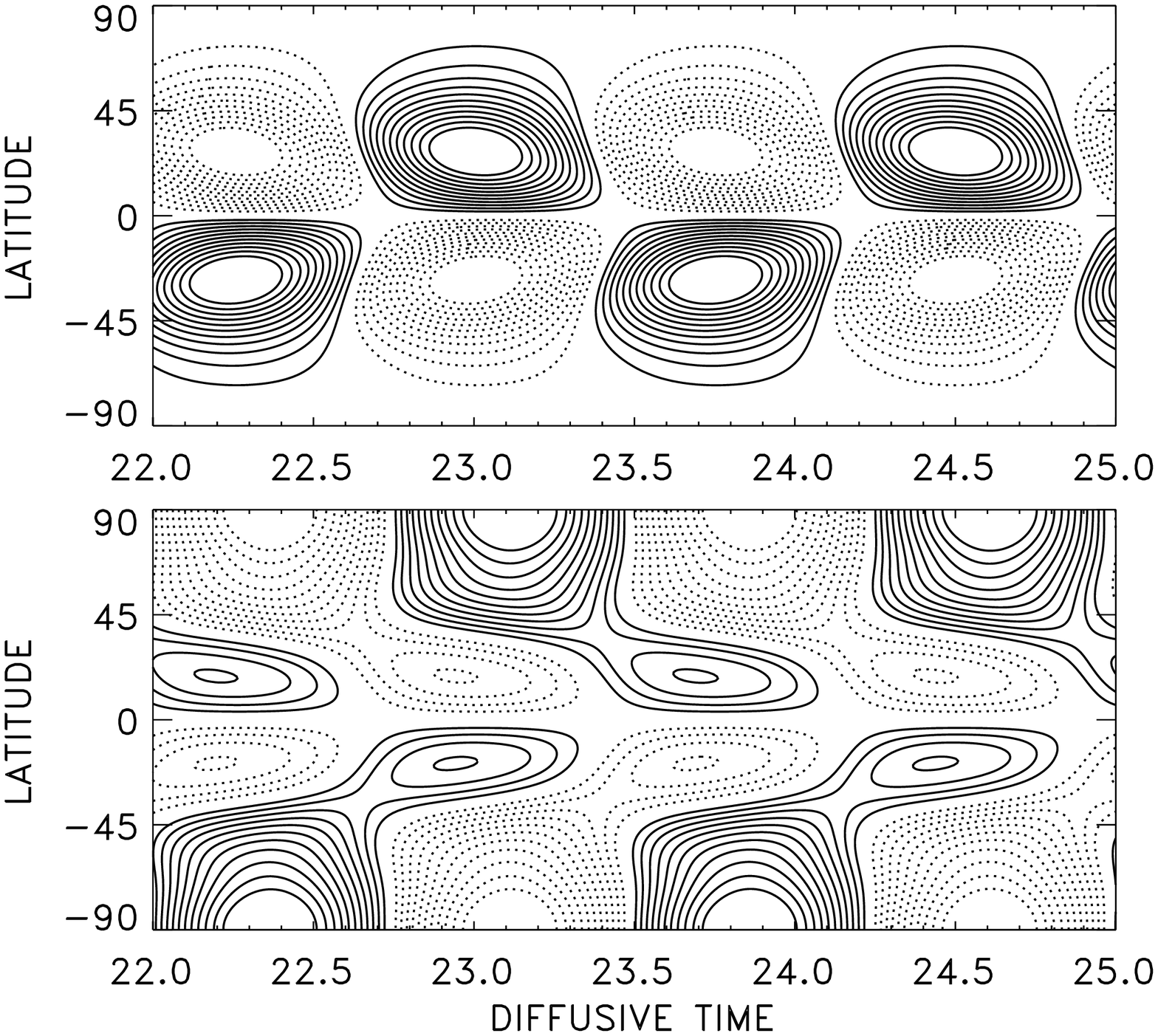}}
 \begin{description}
 \item{\small {\bf Fig.~5.} Isolines of the depth-integrated toroidal
    magnetic field $\cal B$ (top) and the radial magnetic field component
    on the surface (bottom) in latitude–time coordinates. The solid and
    dotted lines indicate positive and negative levels, respectively.
    The time is given in units of the diffusion time $R_\odot^2/\eta_0$.
    The quantity $\cal B$ is defined in (\ref{14}).
    }
 \end{description}
\end{figure}

Figure 5 shows the calculated evolution of the field in
latitude–time coordinates. For comparison with the butterfly diagram
of sunspots, the toroidal field diagrams are usually constructed for
some fixed depth. In the model under consideration, the toroidal
field is concentrated in a thin near-bottom layer, where it strongly
depends on the depth. Therefore, Fig. 5 shows a diagram for the
depth-integrated toroidal field,
 \begin{equation}
    {\cal B}(\theta ) = \sin\theta \int\limits_{x_\mathrm{i}}^1 B(x,\theta )x^2\ \mathrm{d}
    x.
    \label{14}
 \end{equation}
The coefficient $\sin\theta$ и in (\ref{14}) allows for the latitude
dependence of the length of the toroidal field flux tubes (the
probability for the rise of magnetic loops with the formation of
sunspots is assumed to be proportional to the length of the magnetic
flux tube).

The toroidal field diagram in Fig. 5 shows an equatorward drift
during magnetic cycles. Note that the Yoshimura (1975) law is
inapplicable for the nonlocal dynamo and the latitudinal drift is
not related to the rotation inhomogeneity. The radial rotation
inhomogeneity in our model is small and changes its sign depending
on the latitude (Fig. 2).

The polar field in Fig. 5 changes its sign when the toroidal field
reaches its largest strength, i.e., at the magnetic cycle maximum.
The poloidal field diagram in this figure is close to that
constructed by Obridko et al. (2006) from observational data and to
the results by Stenflo (1988).

The presence of calculated toroidal fields at excessively high
latitudes remains the only obvious disagreement of the proposed
model with observations. The sunspot activity on the Sun is limited
to equatorial latitudes, below approximately 30$^\circ$. The
toroidal field diagram in Fig. 5 covers a wider range of latitudes.
It is hoped that this disagreement can be alleviated by taking into
account the meridional flow. Near the base of the convection zone
where the toroidal fields are localized, the meridional flow is
directed equatorward and must lead to the field concentration near
the equator. Another possibility for improving the model is to study
the alpha-effect due to the rise of magnetic loops. Insufficient
knowledge of this effect leads to certain arbitrariness (free
parameters) in the model. At the same time, even in its simplest
present-day version, the model is consistent with observations in a
number of parameters (the cycle period, the equatorial symmetry, the
relation between the amplitudes and signs of the toroidal and
poloidal fields, the equatorial field drift), showing that it is
promising.

\bl
\centerline{ACKNOWLEDGMENTS}
This work was supported by the Russian Foundation
for Basic Research (project nos. 10-02-00148 and 10-02-00391).
\bl
\setlength{\baselineskip}{0.5 truecm}

\centerline{REFERENCES}
\begin{description}
\item Babcock~H.W.: 1961, \apj\ {\bf 133}, 572.
\item Belvedere~G., Kuzanyan~K.M., Sokoloff~D.D.: 2000, \mnras\ {\bf 315},
    778.
\item Brandenburg~A., K\"apyl\"a~P.J.: 2007, New J. Phys. {\bf 9}, 305.
\item Brandenburg~A., Subramanian~K.: 2005, Phys. Rep. {\bf 417}, 1.
\item Brandenburg~A., R\"adler~K.-H., Schrinner~M.: 2008,
    \aa\ {\bf 482}, 739.
\item Brandenburg~A., Kemel~K., Kleeorin~N., Rogachevskii~I.: 2010,
    arXiv:1005.5700.
\item Caligari~P., Moreno-Insertis~F., Sch\"ussler~M.: 1995,
    \apj\ {\bf 441}, 886.
\item Chatterjee~P., Nandy~D., Choudhuri~A.R.: 2004, \aa\ {\bf 427}, 1019.
\item Dasi-Espuig~M., Solanki~S.K., Krivova~N.A. et al.: 2010, \aa\ {\bf 518}, 7.
\item Dikpati~M., Charbonneau~P.: 1999, \apj\ {\bf 518}, 508.
\item Dorch~S.B.F., Nordlund~\AA.: 2001, \aa\ {\bf 365}, 562.
\item Durney~B.R.: 1995, \sp\ {\bf 160}, 213.
\item Gilman~P.A.: 1992, ASP Conf. Series {\bf 27}, 241.
\item Guerrero~G. and de~Gouveia~Dal~Pino~E.M.: 2008, \aa\ {\bf 485}, 267.
\item Howard~R.F.: 1996, Annu. Rev. Astron. Astrophys. {\bf 34}, 75.
\item Ivanova~T.S., Ruzmaikin~A.A.: 1976, Astron. Zh. {\bf 53}, 398
    [Sov. Astron. {\bf 20}, 227 (1976)].
\item Kitchatinov~L.L., R\"udiger~G.: 2008, \an\ {\bf 329}, 372.
\item Krauze~F., R\"adler~K.-H.: 1980, {\sl Mean-Field Electrodynamics
    and Dynamo Theory} (Pergamon, Oxford; Mir, Moscow, 1984).
\item Krivodubskii~V.N.: 1984, Astron. Zh. {\bf 61}, 354 [Sov.
    Astron. {\bf 28}, 205 (1984)].
\item Mason~J., Hughes~D.W., Tobias~S.M.: 2008,
\mnras\ {\bf 391}, 467.
\item Moffat~H.K.: 1978, {\sl Magnetic Field Generation in Electrically
    Conducting Fluids} (Cambridge Univ., London,
    New York; Mir, Moscow, 1980).
\item Nefedov~S.N., Sokoloff~D.D.: 2010, Astron. Rep. {\bf 54}, 247.
\item Obridko~V.N.: 1985, {\sl Sunspots and Activity Complexes}
    (Nauka, Moscow) [in Russian].
\item Obridko~V.N., Sokoloff~D.D., Kuzanyan~K.M., Shelting~B.D.,
    Zakharov~V.G.: 2006, \mnras\ {\bf 365}, 827.
\item Parker~E.N.: 1955, \apj\ {\bf 122}, 293.
\item Parker~E.N.: 1993, \apj\ {\bf 408}, 707.
\item R\"udiger~G., Brandenburg~A.: 1995, \aa\ {\bf 296}, 557.
\item Spence~E.J., Nornberg~M.D., Jacobson~C.M. et al.: 2007,
    Phys. Rev. Lett. {\bf 98}, 4503.
\item Stenflo~J.O.: 1988, Astrophys. Space Sci. {\bf 144}, 321.
\item Vainshtein~S.I., Zel’dovich~Ya.B., Ruzmaikin~A.A.: 1980,
    {\sl Turbulent Dynamo in Astrophysics} (Nauka,
    Moscow) [in Russian].
\item Yoshimura~H.: 1975, \apj\ {\bf 201}, 740.
\item Zel’dovich~Ya.B.: 1956, JETP {\bf 4}, 460.
\end{description}

{\sl Translated by G.~Rudnitskii}
\end{document}